\documentclass[epj]{svjour}

\usepackage{amsmath}
\usepackage{amssymb}
\usepackage{graphicx}
\usepackage{subfigure}
\usepackage{mathrsfs}
\usepackage{xspace}
\usepackage[pdftex,
      colorlinks=true,
      urlcolor=blue,      
      filecolor=blue,      
      linkcolor=blue,       
      citecolor=blue,
      pdfauthor={Georgios Choudalakis, Diego Casadei},
      pdfpagemode=None,
      bookmarksopen=true]{hyperref}

\def\pval{{\ensuremath{p\text{-value}}}\xspace}
\def\pvals{{\ensuremath{p\text{-values}}}\xspace}
\def\zval{{\ensuremath{z\text{-value}}}\xspace}
\def\zvals{{\ensuremath{z\text{-values}}}\xspace}
\def\di{{\ensuremath{\,\text{d}}}\xspace}
\def\B{{\ensuremath{B}}\xspace}
\def\eff{{\ensuremath{\varepsilon}}\xspace}

\begin{document}

\title{Plotting the Differences Between Data and Expectation}

\author{Georgios Choudalakis\inst{1}
 \and Diego Casadei\inst{2}}

\institute{%
  University of Chicago, Enrico Fermi Institute, 
  5640 South Ellis Avenue, Chicago, IL-60637, USA\\
  \email{gchouda@alum.mit.edu}
 \and
  Department of Physics, New York University, 
  4 Washington Place, New York, NY-10003, USA\\
  \email{diego.casadei@cern.ch}
}

\abstract{%
 This article proposes a way to improve the presentation of histograms
 where data are compared to expectation.  Sometimes, it is difficult
 to judge by eye whether the difference between the bin content and
 the theoretical expectation (provided by either a fitting function or
 another histogram) is just due to statistical fluctuations.  More
 importantly, there could be statistically significant deviations
 which are completely invisible in the plot.  We propose to add a
 small inset at the bottom of the plot, in which the statistical
 significance of the deviation observed in each bin is shown.  Even
 though the numerical routines which we developed have only
 illustration purposes, it comes out that they are based on formulae
 which could be used to perform statistical inference in a proper way.
 An implementation of our computation is available at: \url{https://github.com/dcasadei/psde}.
\PACS{
      {02.70.Rr}{General statistical methods}
     } 
} 

\maketitle


\section{Introduction}

 Most analyses compare the observed data to the expectation resulting
 from a theoretical model, like the Standard Model (SM), or some other
 hypothesis, like a best fitting function, or a Monte Carlo simulated
 distribution.  It is common to make histograms in logarithmic scale,
 because their contents span orders of magnitude.  Since differences
 are difficult to see in a logarithmic scale, an inset plot is often
 made at the bottom of the histogram.  In this article, we focus on
 this inset and propose a way to make it intuitive and accurate.  By
 ``intuitive'' we mean that it should make it obvious which bins
 contain an excess of data and which contain a deficit, while
 significant deviations should look more striking than insignificant
 ones.  By ``accurate'' we mean that it should represent the actual
 significance of the deviation in each bin, rather than some
 approximation.

 Below we define statistical significance in two probability models,
 Poisson and binomial. In Section~\ref{sec:progression} we show a
 sequence of presentation options that could be used (and most of them
 have been used), explaining their strengths and limitations.  This
 will motivate, through a series of incremental improvements, our
 final proposal, which is given in Section~\ref{sec:final}.  Finally,
 if the expectation is only known within some uncertainty, this
 uncertainty can be taken into account in the comparison to the data,
 as explained in Section~\ref{sec:syst}.  In this paper, we use the
 \texttt{ROOT} framework \cite{ROOT} to produce the plots and provide
 suggestions for implementing our formulae with this widely used
 software for data analysis and visualization, although our
 recommendations can be easily implemented within other frameworks too.


\subsection{Definition of statistical significance}
\label{sec:signif}

\begin{figure}
  \centering
  \includegraphics[width=0.5\textwidth]{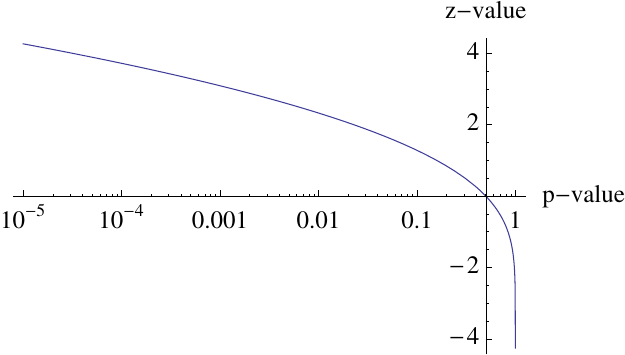}
  \caption{Relationship between \pval and \zval.  \label{fig:pvalzval}}
\end{figure}

 The key notion, which is needed to define the statistical
 significance, is the probability of finding a deviation at least as
 big as the one observed in the data, under the assumption that the
 chosen theoretical model describes our system.  This probability is
 commonly called the ``$p$-value'' and it usually spans several orders
 of magnitude.  For this reason, it is convenient to translate the
 \pval into a ``\zval'', which is the deviation at the right of the
 mean of a Gaussian distribution, expressed in units of standard
 deviations, which would correspond to the same \pval.  The equation
 which maps each \pval to a \zval, and vice versa, is:
\begin{equation}
  \pval = \int_{\zval}^{\infty} \frac{1}{\sqrt{2\pi}} e^{-\frac{x^2}{2}}dx,
\end{equation}
 which is shown in Fig.~\ref{fig:pvalzval}.  In {\tt ROOT}, the \zval
 can be computed in one line of code, using the inverse error
 function:
\begin{equation}
  \nonumber {\tt sqrt(2.)*TMath::ErfInverse(1. - 2.*pvalue)}
\end{equation}

 A $\zval\ge0$ corresponds to a $\pval\le0.5$, and negative \zvals
 correspond to $\pval>0.5$.  Significant deviations are characterized
 by quite small $p$-values, corresponding to \zvals$\ge 3$.
 For example, it is common to refer to a $\pval=2.87\times
 10^{-7}$ as a ``5$\sigma$ effect'', meaning that the corresponding
 \zval is 5.  Such deviations are usually considered very significant
 and a \zval of 5 or more is conventionally required in High Energy
 Physics to claim a discovery \cite{casadeiPHSTAT2011}.  On the other
 hand, a \zval which is less than 1--2 units represents a common
 statistical fluctuation, which is uninteresting.  The fluctuations are
 even more probable (i.e.~less interesting) when the \zval is
 negative.

 The \pval depends on the statistical distribution that the data are
 supposed to follow, the so-called ``probability model''.  The cases
 of Poisson and binomial distributed data are most common in
 experimental physics, and are addressed in this note.


\subsection{Poisson model}
\label{sec:poisson}

 This is the most common case, where event counts are plotted in each
 bin and the number of entries in each bin follows a Poisson
 distribution (e.g.\ Ref.~\cite{dijet}).  If $\B\in\mathbb{R}$ events
 are expected in a bin, the probability of observing $D\in\mathbb{N}$
 events is
\[
  P(D|\B) = \text{Poi}(D|\B) = \frac{\B^D}{D!} \, e^{-\B} 
\]
 ($B$ is also the variance of $D$).  The Poisson
 \pval is
 \begin{equation}
	\pval = 
	\begin{cases}
	  \displaystyle\sum_{n = D}^\infty \frac{\B^n}{n!}\, e^{-\B}
          = 1 - \sum_{n = 0}^{D-1} \frac{\B^n}{n!}\, e^{-\B}
          & , \; D > \B \\
	  \displaystyle\sum_{n = 0}^D \frac{\B^n}{n!}\, e^{-\B}
          & , \; D \le \B \\
	\end{cases}
	\label{eq:pval}
 \end{equation}
 The above sums are simplified thanks to the identity:
 \begin{equation}
    \sum_{n = 0}^{D-1} \frac{\B^n}{n!}\, e^{-\B}
       = \frac{\Gamma(D,\B)}{\Gamma(D)},
 \end{equation}
 where
 \begin{eqnarray}
	\Gamma(D,\B) &=& \int_{\B}^\infty t^{D-1} e^{-t}\; dt, \\
	\Gamma(D) &=&    \int_0^\infty t^{D-1} e^{-t}\; dt.
 \end{eqnarray}

 The ratio between the upper incomplete Gamma function
 $\Gamma(s,x)$ and the Gamma function $\Gamma(s)$ in the
 previous equation is known as the upper regularized Gamma
 function
 \begin{equation}
    Q(s,x)=\frac{\Gamma(s,x)}{\Gamma(s)}=1-P(s,x)
 \end{equation}
 where $P(s,x)$ is the cumulative distribution function for Gamma
 random variables with shape parameter $s$ and scale parameter 1
 \cite{AbramowitzStegun1965}.  In \texttt{ROOT} this function is
 available as
 \[
    Q(s,x) = \texttt{ROOT::Math::inc\_gamma\_c(s,x)},
 \]
 such that
 \begin{equation}
	\pval = 
	\begin{cases}
	   \; 1 - Q(D,\B) = \text{\tt ROOT::Math::inc\_gamma\_c(D,B)}
          & , \; D > \B \\
	   \; Q(D+1,\B) = \text{\tt ROOT::Math::inc\_gamma\_c(D+1,B)}
          & , \; D \le \B \\
	\end{cases}
	\label{eq-pval-root}
 \end{equation}

 
 One should notice that it is not always true that the bin population
 follows the Poisson distribution.  This does not happen when the
 total number of entries in a histogram is not a random variable, in
 which case the events in each bin are distributed accordingly to the
 multinomial distribution, which means that the bins cannot be
 considered statistically independent as we assumed in this section.
 In the extreme case of a fixed total number of entries and only two
 bins, the content of each bin follows the binomial distribution: if
 there is an excess of counts in one bin, with respect to the
 expectation, then the other bin must have a deficit in order to
 preserve the total number of events.


\subsection{Binomial model}
\label{sec:binomial}


 If the plotted quantity is a Bernoulli success rate, then it follows
 a binomial distribution.  A common example would be a trigger
 efficiency ``turn-on curve'' like Fig.~1 in
 Ref.~\cite{jetCrossSection}, where it would be informative to show
 how significantly the observed trigger rate differs from the
 simulation.  Another example is $F_\chi(m_{jj})$ in
 Ref.~\cite{dijetAngular}: if this event rate were significantly
 different than the SM prediction, this could be a sign of certain new
 physics.

 Let us define the following notation:
 \begin{description}
 \item[$n \in \mathbb{N}$:] the number of events initially
       observed in a bin, a subset of which will survive the
       selection;

 \item[$k_{\text{obs}} \in \{0,1,\ldots,n\}$:] the observed number of
       selected events;

 \item[$\eff \in \mathbb{R}$:] the expected success rate,
       i.e.~the selection efficiency, to which we compare the
       observed success rate $k_{\text{obs}}/n$.
 \end{description}
 The binomial distribution gives the probability of observing $k$
 events passing the selection out of the initial $n$ events:
 \begin{equation}
	P(k|n,\eff) = \text{Bi}(k|n,\eff)
	   = \binom{n}{k} \, \eff^k (1-\eff)^{n-k}
 \end{equation}
 The expected number of surviving events is
 $E[k|n,\eff] = n \eff$, with variance
 $V[k|n,\eff] = n \eff (1-\eff)$.
 Finally, the Binomial \pval is defined as
 \begin{equation}
	\pval = 
	\begin{cases}
	  \displaystyle
     \sum_{n = k_{\text{obs}}}^n \binom{n}{k} \,
     \eff^k (1-\eff)^{n-k}
          &  , \;  k_{\text{obs}} \ge n \eff
  \\
	  \displaystyle
     \sum_{n = 0}^{k_{\text{obs}}} \binom{n}{k} \,
     \eff^k (1-\eff)^{n-k}
          &  , \;  k_{\text{obs}} < n  \eff
	\end{cases}
 \end{equation}

 The cumulative distribution function of the binomial model can
 be represented in terms of the regularized incomplete Beta
 function \cite{AbramowitzStegun1965}
\(
  I_x(a,b) = \int_0^x t^{a-1} (1-t)^{b-1} \di t / B(a,b)
\)
 where the Euler's Beta function can be represented in terms of the
 Gamma function as 
\(
  B(a,b) = \Gamma(a) \Gamma(b) / \Gamma(a+b)
\). 
 The regularized incomplete beta function is available in ROOT as 
\[
  I_x(a,b) = \texttt{ROOT::Math::inc\_beta(x,a,b)},
\]
 which makes it possible to implement what follows.

 Let us start with the deficit case $k_{\text{obs}} < n  \eff$,
 for which the \pval coincides with the cumulative distribution
 function:
\begin{equation}
  \pval = P(k\le k_{\text{obs}})
        = I_{1-\eff}(n-k_{\text{obs}}, k_{\text{obs}}+1)
	\; , \quad  k_{\text{obs}} < n \eff \; . 
\end{equation}
 For the case of an excess, we use two identities.  The first is
 $I_x(a,b) = 1 - I_{1-x}(b,a)$ and the second comes directly from the
 binomial distribution and is the starting point of the following
 chain of equations, whereas the previous identity is used in the last
 step:
\[
  P(k \ge k_{\text{obs}}+1) = 1 - P(k\le k_{\text{obs}})
     = 1 -  I_{1-\eff}(n-k_{\text{obs}}, k_{\text{obs}}+1)
     = I_{\eff}(k_{\text{obs}}+1, n-k_{\text{obs}})
\]
 from which we immediately find the \pval for the excess:
\begin{equation}
  \pval = P(k\ge k_{\text{obs}})
        = I_{\eff}(k_{\text{obs}}, n-k_{\text{obs}}-1)
	\; , \quad  k_{\text{obs}} > n \eff
\end{equation}
 which is defined for $k_{\text{obs}} < n$ (in case $k_{\text{obs}}=n$
 one has $P(k\ge n) = P(k=n) = \text{Bi}(n|n,\eff) = \eff^n$).


\subsection{Unfolded data}

 If the histogram is the product of some unfolding procedure, things
 are much less clear, because the bin populations do not follow a
 well-known and generally applicable probability distribution.  In
 addition, the correlations across adjacent bins are unavoidable.  It
 is unclear even if one should try to compute, bin by bin, the
 significance of the difference between unfolded data and theoretical
 expectation.  This could mislead people to think that the bins are
 independent, which is not true.  For this reason, this document does
 not deal further with unfolded spectra.


\section{Some ways to present deviations in Poisson distributed data}
\label{sec:progression}

 The goal of this section is to present a logical progression of
 plotting options, which will lead to the final proposal.  We examine
 the different options in detail because they continue to be used
 despite many people know their drawbacks.  All the examples will use
 the observed (from Monte Carlo simulation) and expected event counts
 of Table~\ref{tab:data} and assume Poisson-distributed bin
 populations, which is the most common case in high-energy physics
 (Sec.~\ref{sec:binomial} explains the treatment of binomial data).
 Two additional features have been introduced in the observed data, an
 excess and a deficit, to make the illustration more interesting.
 In case the data follow neither a Poisson nor a binomial
 distribution, one needs to compute the \pval according to the
 probability distribution which the data are assumed to follow.
 Besides the different \pval definition, the rest of this discussion
 still applies.

\begin{table}
\center \small
  \begin{tabular}{c|rl || c|rl}
    Bin number & Observed & Expected & Bin number & Observed & Expected \\
    \hline \hline
    1 & 121687 &   121688   &      21 & 233 &     226.509    \\ 
    2 & 222014 &   221422   &      22 & 129 &     144.086    \\ 
    3 & 223741 &   223832   &      23 & 90 &      91.4576    \\ 
    4 & 189486 &   190065   &      24 & 51 &      57.9372    \\ 
    5 & 148673 &   148218   &      25 & 53 &      36.6361    \\ 
    6 & 109932 &   109876   &      26 & 20 &      23.1279    \\ 
    7 & 79553 &    78760.1  &      27 & 15 &      14.5779    \\ 
    8 & 56420 &    55119.7  &      28 & 13 &      9.17558    \\ 
    9 & 38771 &    37889.4  &      29 & 3 &       5.76764    \\ 
    10 & 25628 &   25684.7  &      30 & 4 &       3.621      \\ 
    11 & 17309 &   17218.4  &      31 & 4 &       2.2707     \\ 
    12 & 11553 &   11438.1  &      32 & 0 &       1.4224     \\ 
    13 & 7573 &    7540.84  &      33 & 1 &       0.890119   \\ 
    14 & 5024 &    4939.65  &      34 & 0 &       0.556496   \\ 
    15 & 3221 &    3217.98  &      35 & 2 &       0.347608   \\ 
    16 & 1828 &    2086.41  &      36 & 0 &       0.216946   \\ 
    17 & 1213 &    1347.11  &      37 & 0 &       0.135291   \\ 
    18 & 788 &     866.585  &      38 & 0 &       0.0843062  \\ 
    19 & 572 &     555.646  &      39 & 0 &       0.0524979  \\ 
    20 & 362 &     355.233  &      40 & 0 &       0.0326686  \\ 
    \hline \hline
  \end{tabular}
  \caption{Observed events ($D$) and expected events ($\B$), used for
  demonstration. \label{tab:data}} 
\end{table}

\subsection{The $D/\B$ ratio}

 A simple, and very common practice (e.g., Fig.~13 in
 Ref.~\cite{jetCrossSection} and Fig.~1 in Ref.~\cite{cdfDijet}), is
 to plot the ratio of data over expectation ($D/\B$), or subtract 1 to
 plot their relative difference, i.e.\ $(D-\B)/\B$.  As shown in
 Fig.~\ref{fig:ratio}, clearly this is the worst way to compare two
 histograms, especially when the bin contents span several orders of
 magnitude.  It is possible to hide significant discrepancies, as it
 will become obvious.

 The biggest disadvantage of this method is that it does not show the
 statistical significance.  It gives the impression that
 low-population bins are fluctuating more significantly than
 high-population bins, which is not true in this case.  The last
 non-empty bin is off scale, and if we un-zoom the vertical axis to
 include it, then the other features will become even harder to see.
 Although the relative difference cannot be smaller than -1 (for
 $D=0$), it has no upper limit, therefore, the absolute values of
 negative and positive deviations are not comparable.  Bins with $D=0$
 would all be at relative difference $-1$, regardless of $\B$, and are
 typically not shown.

 The only advantage of this method is that excesses of data appear
 above 0, and deficits below, which is intuitive.

 Although this method is inappropriate to convey the statistical
 significance of deviations, it might still be used when a statistical
 comparison is not desired.  For example, one may wish to show simply
 that the data lie within 5\% from the expectation, even if a 5\%
 deviation would be statistically significant in regions with many
 events and insignificant in regions with just a few.

\subsection{The $(D-\B)/\sqrt{\B}$ approximation}

\begin{figure}
\begin{minipage}[t]{0.49\textwidth}
  \centering
  \includegraphics[width=\textwidth]{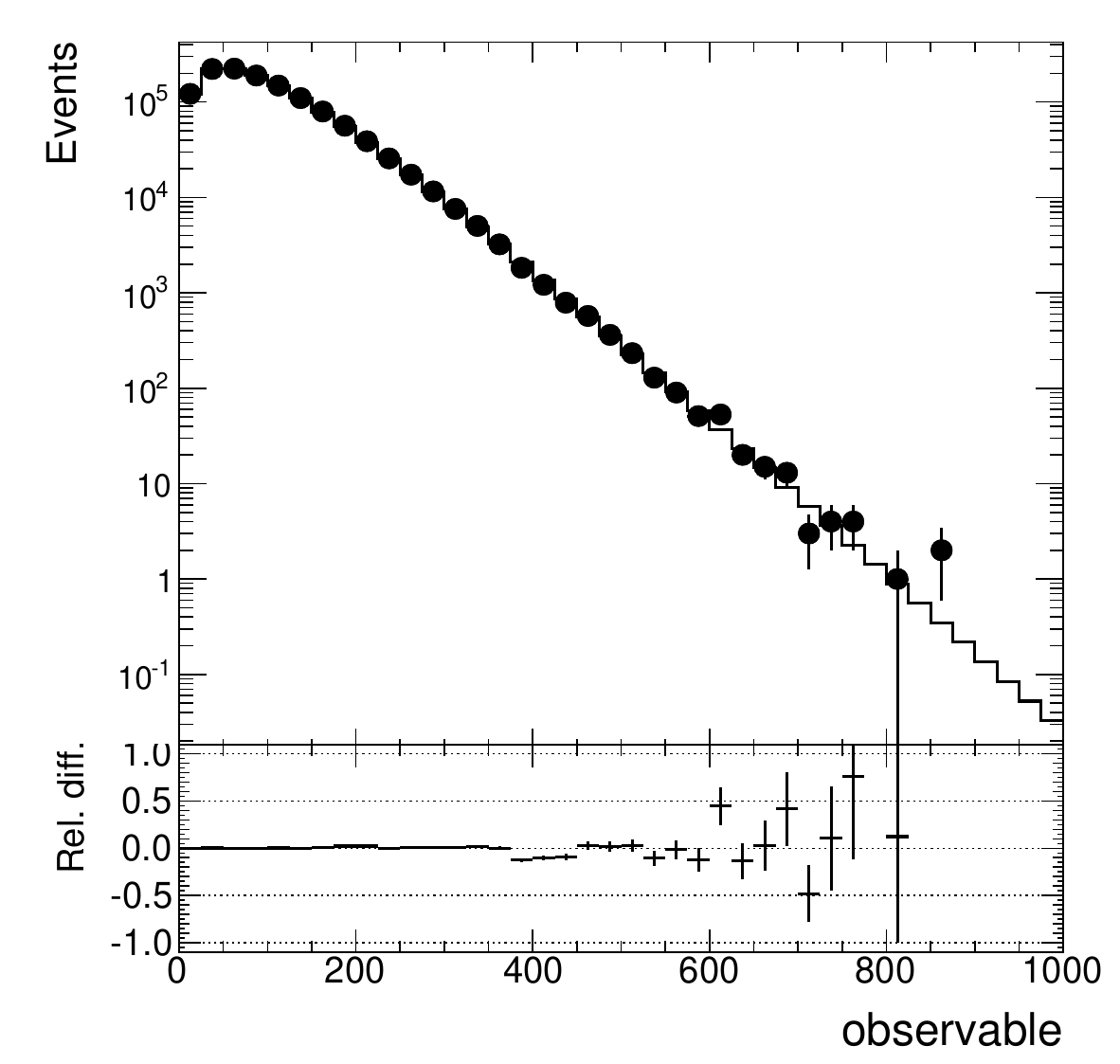}
  \caption{The relative difference between data ($D$) and expectation
  ($\B$): $(D-\B)/\B$ is not a good indicator of the statistical
  significance of the observed deviations.}
  \label{fig:ratio}
\end{minipage}\hfill
\begin{minipage}[t]{0.49\textwidth}
  \centering
  \includegraphics[width=\textwidth]{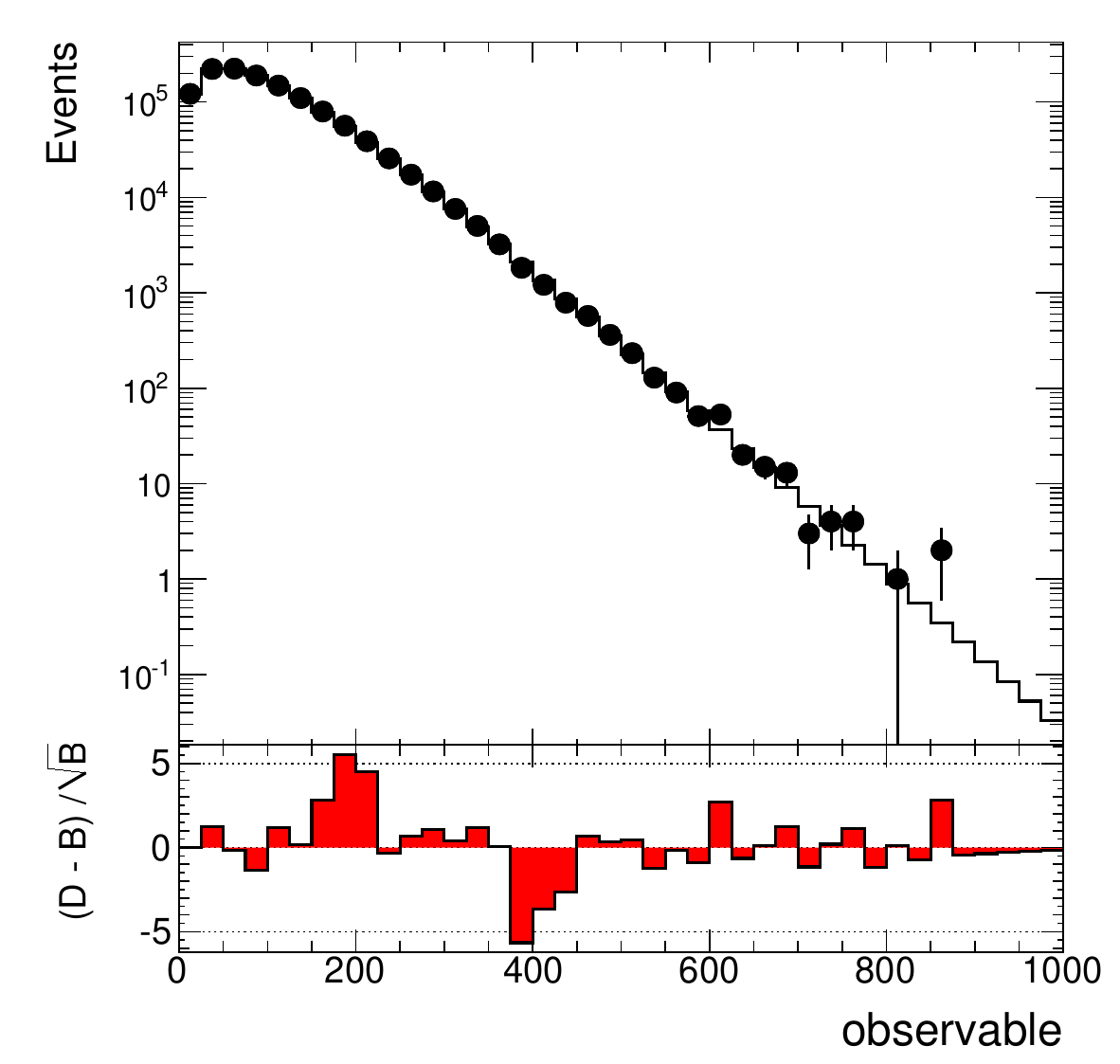}
  \caption{The quantity $(D-\B)/\sqrt{\B}$ approximates the
  statistical significance in bins with large population.  However,
  this approximation fails when there are only a few entries.}
  \label{fig:sqrtB}
\end{minipage}%
\end{figure}

 It is well-known that, for large $\B$, the Poisson distribution with
 parameter $\B$ is approximated well by a Gaussian of mean $\B$ and
 standard deviation $\sqrt{\B}$.  This leads to the approximation of
 the Poisson \zval with $\frac{D-\B}{\sqrt{\B}}$, which is plotted in
 Fig.~\ref{fig:sqrtB}.  For example, Ref.~\cite{dijetOld} contains such a plot.

 An advantage of this approach is that we see clearly two significant
 features: an excess around 200 and a deficit around 400.
 Intuitively, excesses point ``up'' and deficits ``down''.  Indeed,
 the quantity shown, in bins with large expectation, is a good
 approximation of the actual Poisson \zval.

 The main disadvantage is that, in low-population bins,
 $\frac{D-\B}{\sqrt{\B}}$ is not a good approximation of the true
 \zval.  We will see soon that the last bin with $D=2$ is not as
 significant as it appears in Fig.~\ref{fig:sqrtB}.  Similarly, the
 empty bins in the end of the spectrum do not actually have deficits of
 \zval between 0 and 1, as is indicated by the small negative bars.
 Such \zvals would correspond to $\pvals < 0.5$, but, as we are about
 to show, these bins contain deficits of $\pval > 0.5$, which the
 approximation in Fig.~\ref{fig:sqrtB} fails to describe.

\subsection{Plotting signed \zvals}
\label{sec:signed}

\begin{figure}
\begin{minipage}[t]{0.49\textwidth}
  \centering
  \includegraphics[width=\textwidth]{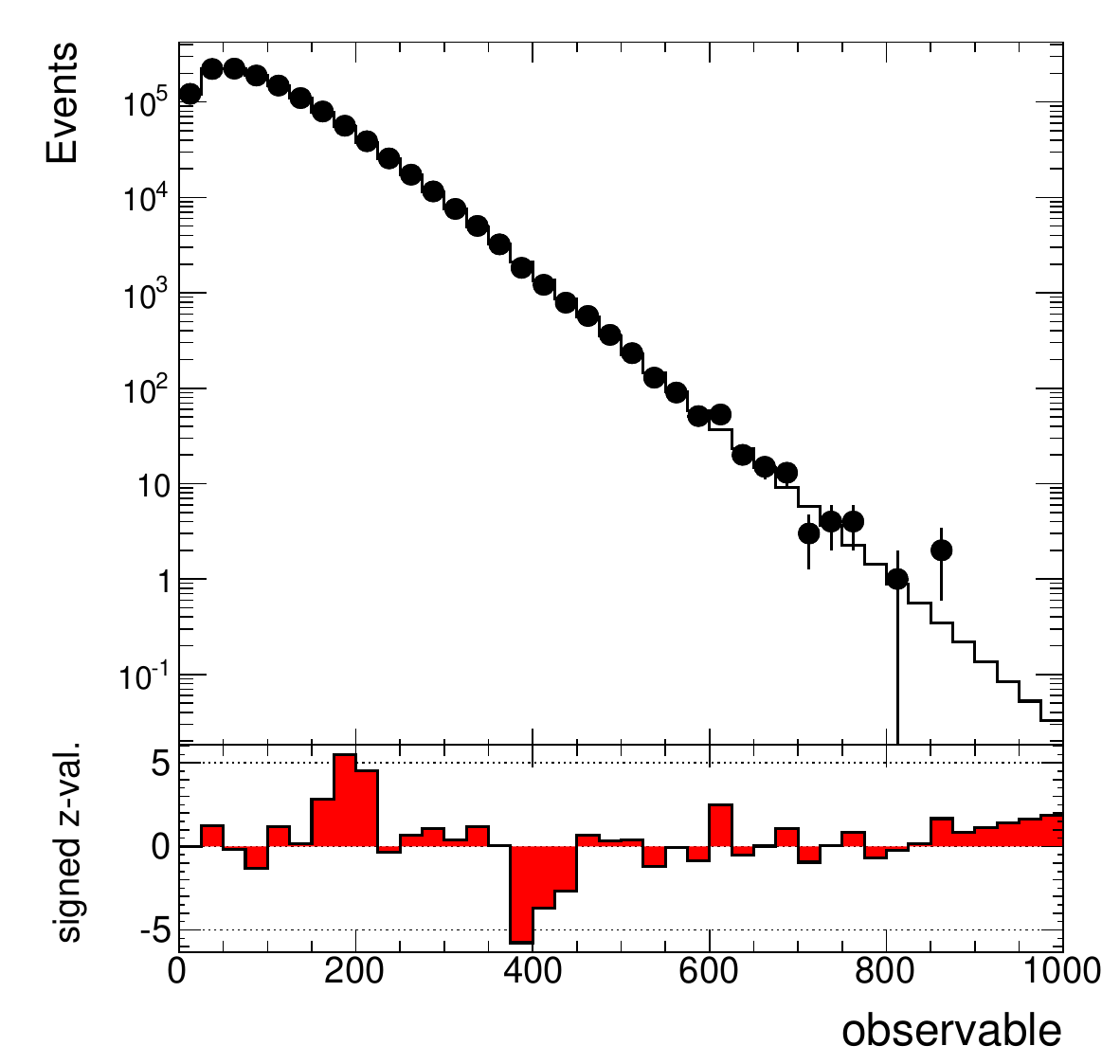}
  \caption{Showing the \zval corresponding to the \pval in
  each bin. In bins with $D<\B$, the $-$\zval is plotted, to distinguish excesses from deficits.}
  \label{fig:signedZval}
\end{minipage}\hfill
\begin{minipage}[t]{0.49\textwidth}
  \centering
  \includegraphics[width=\textwidth]{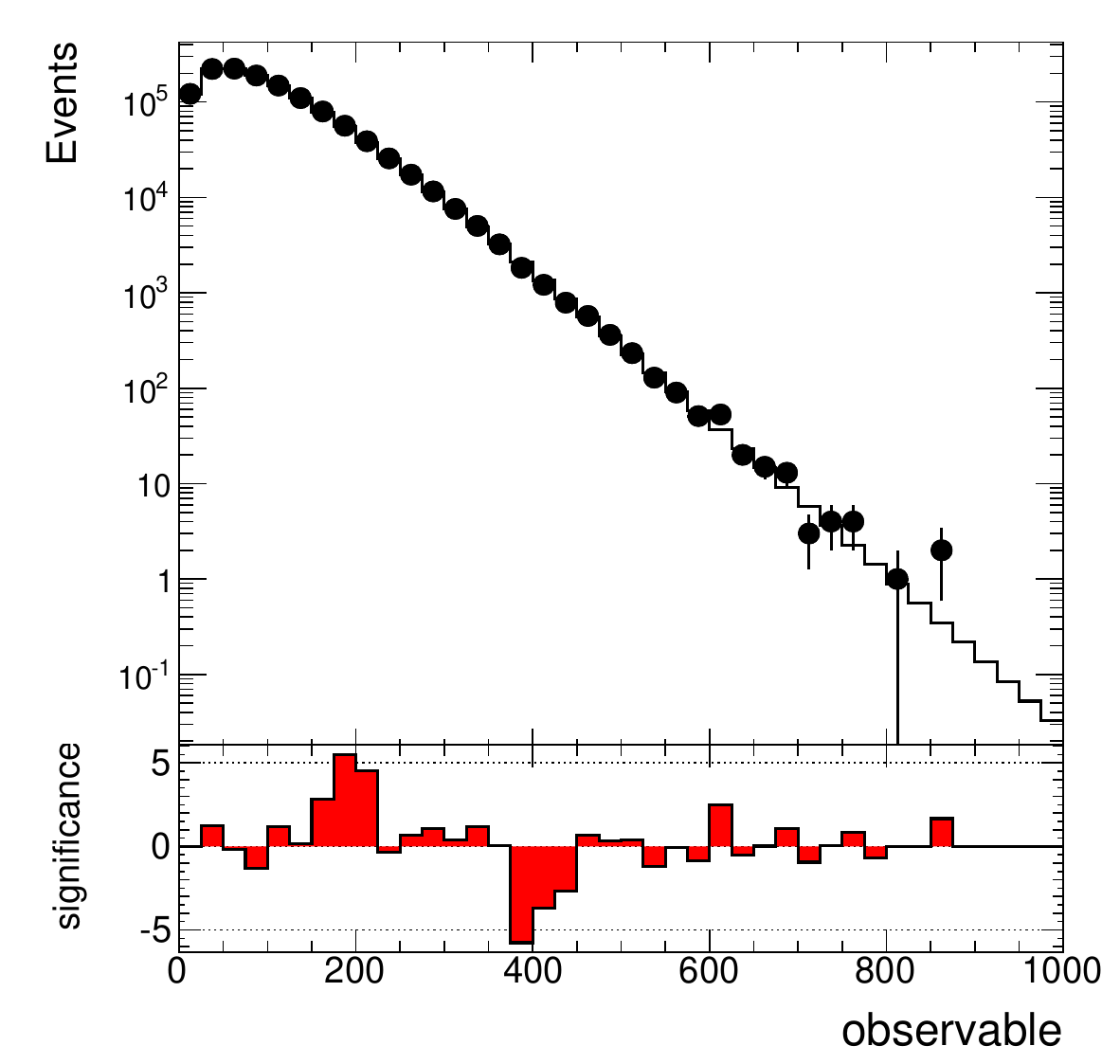}
  \caption{The final proposal.  Plotting the \zval only for bins with
  $\pval < 0.5$, with positive sign for excesses and negative for deficits.}
  \label{fig:signedZvalNonNeg}
\end{minipage}%
\end{figure}

 Instead of approximating the \zval with $\frac{D-\B}{\sqrt{\B}}$, it
 would be better to plot the exact \zval in each bin.  If a bin
 contains a significant excess of data (with $\pval < 0.5$), it will
 have a $\zval > 0$ which, when plotted, will give correctly the
 impression of an excess.  If a bin contains a significant deficit of
 data (with $\pval < 0.5$), it will have a $\zval > 0$ as well, so, to
 avoid giving the impression of an excess in bins with deficits,
 instead of \zval we plot $-\zval$ in these bins.  The result is in
 Fig.~\ref{fig:signedZval}.

 This is a more accurate representation of significance than
 $\frac{D-B}{\sqrt{B}}$, but there is still something confusing in
 bins with low statistics.  We know that the last few bins contain
 deficits ($D=0$), however, they appear in Fig.~\ref{fig:signedZval}
 as if they were excesses.  This happens because they originally had
 negative \zvals, indicating very insignificant deficits, so, with the
 sign-flipping they appear like excesses. The less significant the
 deficit, the larger the \pval, and the more negative the $\zval$,
 which gives the impression of an even larger excess when plotting
 $-$\zval.  The same problem, but in the opposite direction, is
 observed in bin 33, around 800.  It contains a very insignificant
 excess with $\zval < 0$, and since it is an excess this $\zval$ is
 plotted without changing sign.  The result points ``down'', giving
 wrongly the impression of a deficit.

\subsection{The final proposal: plot signed \zvals only if $\pval<0.5$}
\label{sec:final}

 In conclusion, our proposal is to plot \zvals as in
 section~\ref{sec:signed}, which are negative in bins with data
 deficits, with the exception of bins with $\pval > 0.5$, whose
 significance should be simply not shown.  The result is shown in
 Fig.~\ref{fig:signedZvalNonNeg}.

 The justification for not showing the significance of bins where
 $\pval>0.5$ is that such bins agree perfectly with expectation:
 such deviations are totally uninteresting.  In addition, trying to
 present their \zvals could be misleading, as explained in
 section~\ref{sec:signed}.

 The advantages of this proposal are the following: 
\begin{itemize}
\item The \zval is accurate, unlike the  $(D-\B)/\sqrt{\B}$ approximation.

\item Positive values represent excesses of data over the expectation,
  and negative values correspond to deficits.

\item No feature is hidden if it is worth showing, namely if it has a
  $\pval < 0.5$.

\item Bins with high and low statistics (like bin 35 with $D=2$ and
  bin 32 with $D=0$) are treated in the same way.

\item If the significance is not shown for a bin, it means that there
  is no relevant deviation from the expectation.

\item What we propose is easy to implement, using existing {\tt ROOT}
  functions as we show in sections~\ref{sec:signif}, \ref{sec:poisson}
  and \ref{sec:binomial}.
\end{itemize}


\section{Including the theoretical uncertainty}
\label{sec:syst}

 Any theoretical uncertainty in the reference value will affect the
 significance of the observation.  For example, in case of a Poisson
 process with parameter \B, a non-zero uncertainty on \B\ will
 decrease the significance of the difference between $D$ and $\B$.
 Because realistic use cases always involve some (perhaps small)
 theoretical uncertainty, it is important to study its effects on our
 estimate of the significance of the result.  A full treatment would
 require to solve an inference problem.  However, here we limit
 ourselves to the aspects which are relevant when making plots, and
 consider only the total theoretical uncertainty, without addressing
 all possible components which may contribute to it.

 Of course, in some cases it makes sense not to show the effect of
 theoretical uncertainties.  For example, this can be done if the
 theoretical uncertainty is negligible compared to the statistical
 uncertainty, or if the deviations without theoretical uncertainty are
 already not significant enough to deserve further investigation.  On
 the other hand, before claiming than an excess is really statistical
 significant one should always check what happens when including the
 total uncertainty on the expectation.

\begin{figure}
\begin{minipage}[t]{0.49\textwidth}
  \centering
  \includegraphics[width=\textwidth]{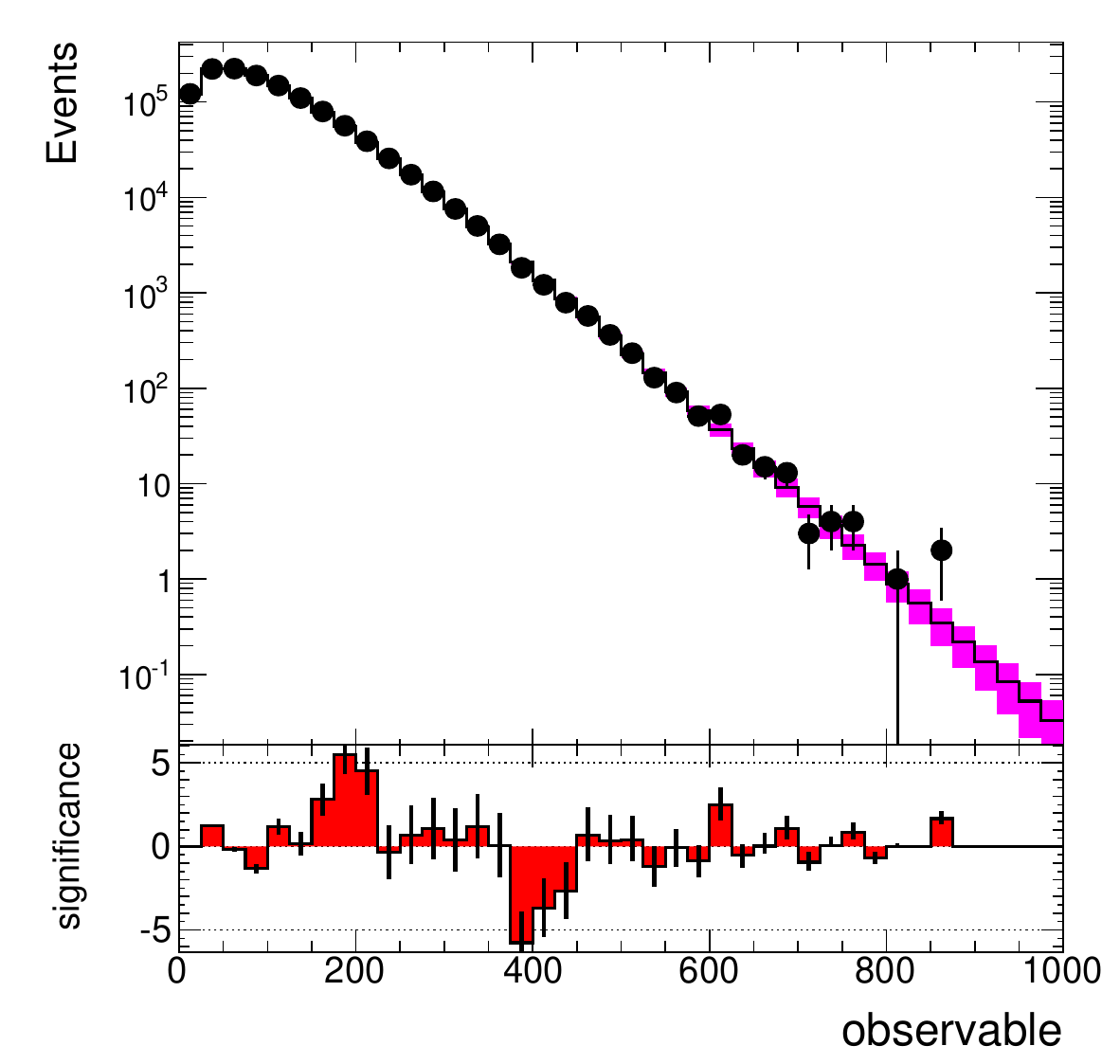}
  \caption{The error bars in the inset indicate the range by which the
  significance varies when recomputed after shifting the expectation
  by $\pm$1 standard deviation, corresponding to the error boxes shown
  in the main plot.  This is \emph{not} the correct way of accounting
  for theoretical uncertainties.}
  \label{fig:errorBars}
\end{minipage}\hfill
\begin{minipage}[t]{0.49\textwidth}
  \centering
  \includegraphics[width=\textwidth]{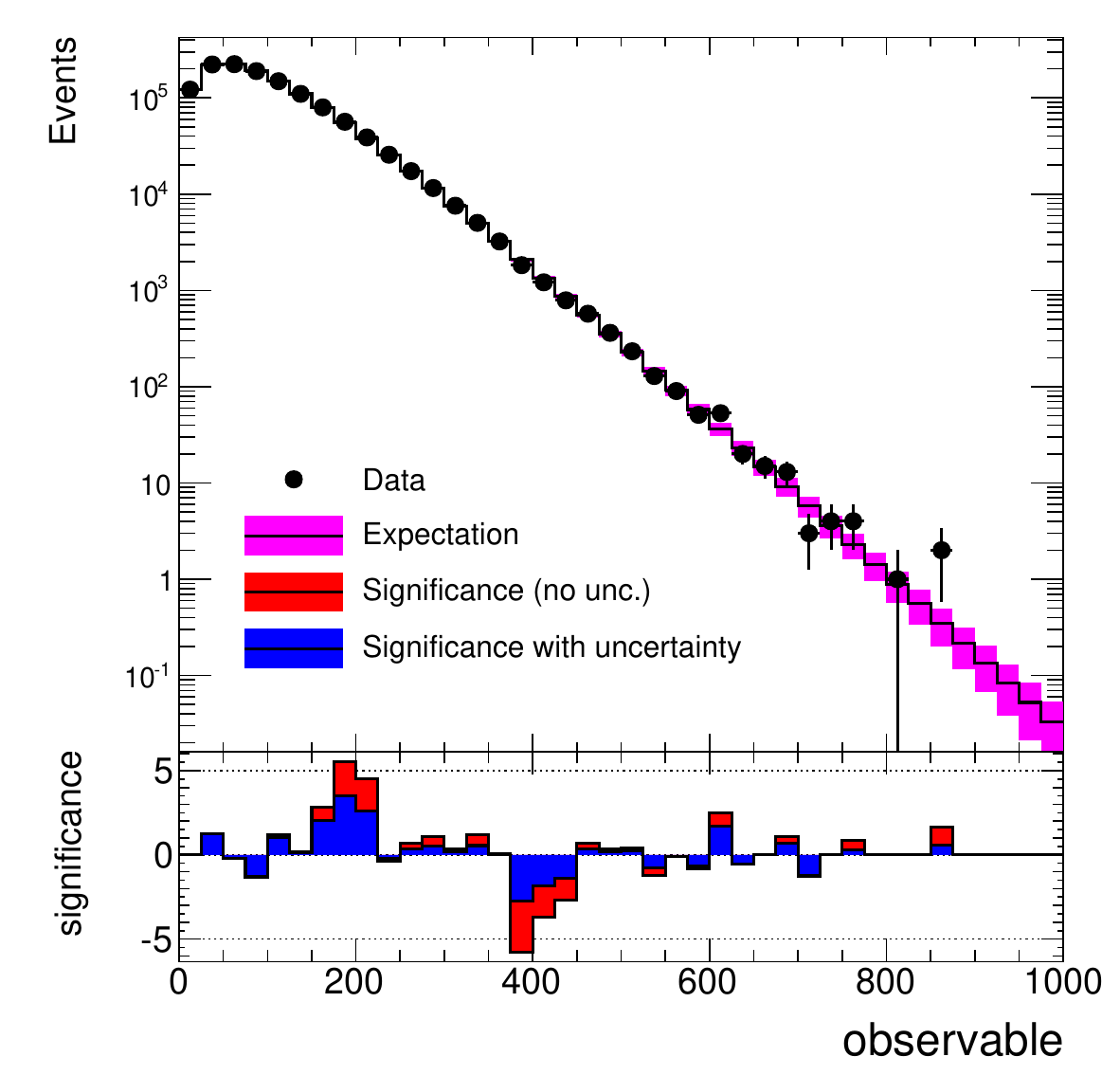}
  \caption{The inset shows the significance computed by neglecting the
  theoretical uncertainty on the expectation (red histogram) and by
  including it (blue histogram).  Including additional sources of
  uncertainty decreases the significance. }
  \label{fig:convolution}
\end{minipage}%
\end{figure}

 At first sight, a simple procedure to warn the reader about the
 effects of the theoretical uncertainty on \B\ would seem to plot the
 significance shown in Fig.~\ref{fig:signedZvalNonNeg} with ``error
 bars''.  The upper and lower deviations with respect to the
 significance computed above could correspond to the result of
 recomputing the significance with the parameter \B\ shifted by $\pm1$
 standard deviation (assuming symmetric uncertainty).  An example of
 this is in Fig.~\ref{fig:errorBars}, where the relative uncertainty
 in bin $i$ has been set to $10^{-5}\cdot i^3$.  The problem with this
 approach is that it misses the fundamental point: \emph{any
 additional uncertainty will decrease the significance of the observed
 deviation}.

 This can be shown with a formal approach, which would correspond to a
 Bayesian treatment of the uncertainties.  It consists of computing
 the \pval with the marginal model obtained after integration over the
 parameter, whose prior describes the experimenter's degree of belief
 on the allowed range of values.  This is explained in detail in the
 next two sections for the Poisson and binomial models, and it is
 illustrated by Fig.~\ref{fig:convolution}, which should be compared
 to Fig.~\ref{fig:errorBars}.

\subsection{Poisson model with uncertain parameter}

 In any bin the probability of observing $n$ events is given by the
 Poisson distribution with parameter $Y>0$ (the expected number of
 events), which is not certain.  One assumes to know its best estimate
 (which we take to be the expectation $B$ of $Y$) and ``uncertainty''
 (which we take to be the square root $S$ of the variance of $Y$).  In
 most cases $B$ and $S$ are the result of some auxiliary measurement.
 We choose to model the uncertainty on the Poisson paramter with a
 function belonging to the Gamma family, which is the conjugate family
 for the Poisson model.  This allows to perform the integration
 analytically.  A unique Gamma density corresponds to the pair
 $(B,S)$.  If we write the Gamma density as
\begin{equation}
 \text{Ga}(x|a,b) = \frac{b^a}{\Gamma(a)} \,
                    x^{a-1} \, e^{-bx}
\end{equation}
 the expectation and variance are
\[
 E = \frac{a}{b} = B
 \quad \text{and} \quad
 V = \frac{a}{b^2} = S^2
\]
 from which we obtain the Gamma parameters
\[
 a = \frac{B^2}{S^2} \quad \text{and} \quad
 b = \frac{B}{S^2} \; .
\]

 In order to evaluate the probability of observing $n$ events, when
 the expected yield is described by the Gamma density above, we make
 use of the marginal model which is obtained by integrating over all
 possible values of $Y$.  The result is the Poisson-Gamma mixture
 (also known as negative binomial)
\begin{equation}\label{eq-pois-marg}
 P(n|a,b) = \int_0^{\infty} \text{Poi}(n|Y) \,
            \text{Ga}(Y|a,b) \di Y
          = \frac{b^a}{\Gamma(a)} \,
            \frac{\Gamma(n+a)}{n! \, (1+b)^{n+a}}
\end{equation}
 (the Gamma function is available in \texttt{ROOT} as $\Gamma(x)=$
 \texttt{ROOT::Math::tgamma(x)}).  We now proceed as above, by defining
 the $p$-value and $z$-value\footnote{Defined as in eq.~\ref{eq:14} and
   \ref{eq:16}, the \zval is automatically negative for deficits.} in
 terms of the marginal model (\ref{eq-pois-marg}):
\begin{equation}
 \label{eq:14}
 \pval = 
 \begin{cases}
   \displaystyle
       P(n \ge n_{\text{obs}})
       = 1 - \sum_{n=0}^{n_{\text{obs}}-1} P(n|a,b)
       = \int_{\zval}^{\infty} \mathcal{N}(x;0,1) \di x
   &  , \;  n_{\text{obs}} > B
       \\
   \displaystyle
       P(n \le n_{\text{obs}})
       = \sum_{n=0}^{n_{\text{obs}}} P(n|a,b)
       = \int_{-\infty}^{\zval} \mathcal{N}(x;0,1) \di x
   &  , \;  n_{\text{obs}} \le B
 \end{cases}
\end{equation}
 (some care needs to be taken when summing over the marginal model
 (\ref{eq-pois-marg}): it is recommended to make use of the recursive
 relations which allow to compute the $n$-th term starting from the
 term $n-1$).  Finally, we plot the \zval\ only if $p\le0.5$ as
 explained above, to obtain the results shown in
 Fig.~\ref{fig:convolution}.

 Often the theoretical uncertainty is modeled with a Gaussian
 distribution, instead of the Gamma density used here.  Provided that
 the part of the Gaussian which extends to negative (unphysical)
 values is negligible, one obtains a result (by numerical integration)
 which is practically identical to what we obtained above (analytically) using Gamma.
 However, in cases where the negative Gaussian tail is non-negligible, the Gaussian 
 needs to be truncated at zero.  This makes its mean parameter different from its
 actual mean, and its width parameter different from its actual standard
 deviation.  In addition, the truncated Gaussian gives non-null
 probability to the value of zero (see Ref.~\cite{cousins0702156} for
 additional comments).  For these reasons, a truncated Gaussian is a
 poor choice for the probability model. It is highly recommended to use a
 density function with the correct domain, as the Gamma density, which
 has the additional advantage of making the integration possible analytically.  

\subsection{Binomial model with uncertain parameter}

 Let the observed ratio be $k_{\text{obs}}/n$ in a bin for which we
 expect a binomial distribution with parameter $\eff\in[0,1]$ which is
 uncertain.  We choose to model our knowledge about \eff\ by means of
 a Beta distribution with known parameters $a,b$.  The Beta functions
 constitute the conjugate family for the binomial model
 \cite{casadei0908.0130}, which allows us to perform the integration
 analytically.  The parameters $a,b$ can be found by with the method
 of moments or numerically.  We assume that they have been determined
 in order to best represent the result of some auxiliary measurement.
 Then the probability of observing $k$ events passing the selection
 out of the initial $n$ events is given by the marginal model
\[
 P(k|n,a,b) = \int_0^1 \binom{n}{k} x^k (1-x)^{n-k}
               \frac{x^{a-1} (1-x)^{b-1}}{B(a,b)} \di x
	    = \frac{\binom{n}{k}}{B(a,b)} \int_0^1
               x^{k+a-1} (1-x)^{n-k+b-1} \di x 
\]
 where the last integral is the definition of the Euler's Beta
 function $B(k+a,n-k+b)$.  Hence
\begin{equation}\label{eq-binom-marg}
 P(k|n,a,b) =  \binom{n}{k} \, \frac{B(k+a,n-k+b)}{B(a,b)}
\end{equation}
 In \texttt{ROOT}, the Beta function $B(x,y) = \Gamma(x) \Gamma(y) /
 \Gamma(x+y)$ is available as
\[
  B(x,y) = \text{\texttt{ROOT::Math::beta(x,y)}}
\]
 which makes the marginal model (\ref{eq-binom-marg}) easy to compute,
 although summing over it is better achieved if the recurrence
 relations which connect $P(k|n,a,b)$ to $P(k-1|n,a,b)$ are exploited.

 The $p$-value and $z$-value are then defined in terms of the marginal
 model (\ref{eq-binom-marg}) as
\begin{equation}
 \label{eq:16}
 \pval = 
 \begin{cases}
   \displaystyle
       P(k\ge k_{\text{obs}})
       = \sum_{k=k_{\text{obs}}}^{n} P(k|n,a,b)
       = \int_{\zval}^{\infty} \mathcal{N}(x;0,1) \di x
   &  , \;  k_{\text{obs}} \ge  n \eff
       \\
   \displaystyle
       P(k\le k_{\text{obs}})
       = \sum_{k=0}^{k_{\text{obs}}} P(k|n,a,b)
       = \int_{-\infty}^{\zval} \mathcal{N}(x;0,1) \di x
   &  , \;  k_{\text{obs}} < n  \eff
 \end{cases}
\end{equation}
 Finally, one plots only the $z$-values which correspond to $p\ge0.5$
 as explained in the previous sections.

 Similarly to the Poisson case, the choice of a Gaussian distribution
 to model the theoretical uncertainty is problematic, because in the binomial case 
 the domain is limited to the interval $[0,1]$, so the Gaussian may have to be
 truncated on both sides.  The Beta density is strongly recommended
 because it has the correct domain and it allows for analytic
 integration.  In cases where the Gaussian distribution does not span much below 0 or above 1,
 it is practically indistinguishable from the Beta distribution with the same mean and standard deviation.

\section{Summary}

 In this article, we propose an improved way of plotting the
 difference between data and expectation, based on the significance of
 the difference in each bin.  We have shown than one can obtain an
 accurate plot of the statistical significance of the deviation of the
 bin contents from the expectation, obtaining at the same time an
 intuitive picture of the relevant deficits and excesses.  This is
 achieved by computing the exact \pval and, when its value is smaller
 that 50\% probability, by mapping it into the \zval which gives the
 deviation in units of Gaussian standard deviations.  The sign of
 \zvals is always positive for excesses and negative for deficits,
 making it trivial to decode the result at first sight.

 When there are important deviations, whose \zvals are above 3 or 4,
 it is fundamental to check what happens by including the total
 uncertainty on the expectation.  This always\footnote{After publication we realized that this statement is not {\em always} true. When the significance without extra uncertainties is large, then indeed adding extra uncertainty lowers the significance. But if the significance is small without considering additional uncertainties, it is actually possible that it increases when additional uncertainties are convoluted. This is a well-known effect which can be explained, qualitatively, in this way: If without extra uncertainties the data agree (almost) perfectly with expectation, then convoluting alternative possibilities could make the agreement worse, since these alternatives cannot possibly entail better than perfect agreement.} lowers the actual
 significance and provides a result which is closer to what one would
 have obtained by performing a proper statistical inference.  Here the
 focus is only on methods which can improve illustrations.  However,
 it is important to notice that our formulae for the \pvals in the
 presence of uncertainty on the expectation coincide with the
 solutions $p_{\Gamma}$ for the Poisson case and $p_{\text{Bi}}$ for
 the binomial case, computed by Cousins, Linnemann, and Tucker
 \cite{cousins0702156} using a completely different approach.  These
 authors have performed frequentist evaluations of the performance of
 such estimators and concluded that they are optimal for statistical
 inference.  We repeat that we do not focus on statistical inference here, 
 but this agreement between completely independent derivations makes us more
 confident about the validity of our proposal.

\section*{Acknowledgements}

 We thank Alex Read, Glen Cowan, and Kyle Cranmer for their feedback.

\bibliographystyle{plain}

%

\end{document}